\documentclass{iopart}

\usepackage{iopams}
\usepackage{graphicx}
\usepackage{xcolor}
\usepackage[square,sort&compress,numbers]{natbib}
\usepackage[colorlinks,linkcolor=blue!80!black,urlcolor=purple!50!black,citecolor=green!50!black]{hyperref}

\def\email#1{\ead{\href{mailto:#1}{#1}}}

\begin{document}

\title{Generation of arbitrary all-photonic graph states from quantum emitters}
\author{Antonio Russo}
\email{aerusso@ucla.edu}
\author{Edwin Barnes}
\email{efbarnes@vt.edu}
\author{Sophia E. Economou}
\email{economou@vt.edu}
\address{Department of Physics, Virginia Polytechnic Institute and State University, Blacksburg, Virginia 24061, USA}
\date{\today}

\def\eqnref#1{Eq.~(\ref{#1})}
\def\eqnsref#1#2{Eqs.~(\ref{#1}) and (\ref{#2})}
\def\figref#1{Fig.~\ref{#1}}
\def\tableref#1{Table~\ref{#1}}
\def\ket#1{\left|#1\right\rangle}
\def\bra#1{\left\langle#1\right|}
\def\braket#1#2{\left\langle#1\middle|#2\right\rangle}
\def\units#1{\,\mathrm{#1}}
\def\CZ{\mathord{\textsc{c}\hspace{-0.15pt}\textsc{z}\,}}
\def\SWAP{\mathord{\,\textsc{s}\hspace{-0.5pt}\textsc{w}\hspace{-0.5pt}\hspace{-1.15pt}\textsc{a}\hspace{-0.75pt}\textsc{p}\,}}

\newenvironment{cmatrix}{\left(\begin{array}{*{4}{r}}}{\end{array}\right)}

\begin{abstract}
We present protocols to generate arbitrary photonic graph states from quantum emitters that are in principle deterministic.  We focus primarily on two-dimensional cluster states of arbitrary size due to their importance for measurement-based quantum computing.  Our protocols for these and many other types of two-dimensional graph states require a linear array of emitters in which each emitter can be controllably pumped, rotated about certain axes, and entangled with its nearest neighbors.  We show that an error on one emitter produces a localized region of errors in the resulting graph state, where the size of the region is determined by the coordination number of the graph.  We describe how these protocols can be implemented for different types of emitters, including trapped ions, quantum dots, and nitrogen-vacancy centers in diamond.

\vspace{1em}\noindent DOI: \href{https://doi.org/10.1088/1367-2630/ab193d}{10.1088/1367-2630/ab193d}
\end{abstract}
\maketitle

\section{Introduction}
\begin{figure}
\centering
\includegraphics{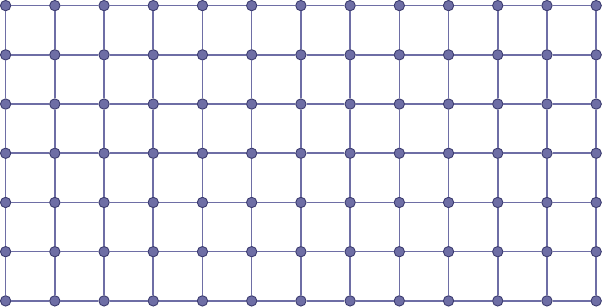}
\caption{Two-dimensional cluster state.  Each vertex represents a qubit, and each edge represents entanglement between the qubits it connects.  This state can serve as a resource for universal measurement-based quantum computation.  \label{fig:2dcluster}}
\end{figure}

Quantum information science holds the promise of providing novel capabilities and computational speedups for a host of problems related to computer science \cite{PhysRevLett.79.325}, secure communication \cite{Shor_1997,Gisin_NatPhoton07,Scarani_RMP09,Ursin_NatPhys07,Liao_Nature17}, the simulation of physical systems \cite{Feynman_1982,Deutsch_1985}, and distributed computation \cite{Beals_PRSA13,VanMeter_Computer16,Sheikholeslami_IEEE16,Fitzsimons_npjQI17}.  One quantum computing paradigm is measurement-based computing \cite{PhysRevA.68.022312,Raussendorf_2012,Nielsen_2006} with a highly entangled \emph{cluster state} (\figref{fig:2dcluster}).  The appeal of this approach is that once a two-dimensional cluster state is created, only local measurements of each qubit are required to perform arbitrary quantum computation.  If the qubits are photons, this corresponds only to waveplate operations and polarized photon counters.

Cluster states have been experimentally demonstrated using trapped ions \cite{Mandel_2003,Lanyon_2013}, squeezed states of light \cite{Yokoyama_2013,PhysRevLett.101.130501,PhysRevLett.112.120505}, superconducting qubits \cite{1811.02292}, and photons \cite{Walther_2005,Tokunaga_2008,Schwartz_2016,PhysRevLett.104.020501}.  The most common approach to constructing multiphotonic entanglement begins with Bell pairs of photons, generated by parametric down conversion, and builds up entanglement pairwise through two-photon interference and measurement (`fusion' gates) \cite{PhysRevLett.95.010501}.  This mechanism is inherently probabilistic, generating two-photon entanglement 50\% of the time, a fact that makes scaling up to larger graph states challenging, especially two- or three-dimensional cluster states.  As a result, the size of photonic graph states generated to date with this approach has been limited to about 10 photons \cite{Gao_2010}.  One-dimensional cluster states have also been created by pumping entangled photons from quantum dot emitters \cite{Schwartz_2016} using a pulsed generation protocol.  This approach coherently drives a quantum dot into an excited state manifold, after which it then relaxes back to a ground state or metastable manifold via spontaneous emission, creating photons entangled with the emitter.  This is an {\it in-principle} deterministic approach in the sense that no probabilistic fusion gates are needed.  However, experimental inefficiencies in photon collection remain a significant obstacle to completely deterministic state generation.

A generalization of the deterministic pulsed generation protocol produces $2\times m$ ``ladder'' photonic cluster states \cite{PhysRevLett.105.093601}.  In this scheme, two quantum dot spins are acted on by a two-qubit controlled phase gate, creating entanglement.  When both dots are then pumped to produce a pair of photons, these photons are entangled with each other.  If this process is repeated multiple times with a periodicity that allows for free, single-qubit precession by an angle of $\pi/2$ between each entangling operation and pumping process, then the resulting state is a $2\times m$ cluster state, where $m$ is the number of cycles.  Schemes to create ladder-type cluster states using trapped ions have also been proposed \cite{Wunderlich_2009}.  However, for universal quantum computation, a larger two-dimensional grid ($n$ by $m$, with $n,m\gg2$) is needed;  moreover, fault-tolerant universal computation requires a three-dimensional grid \cite{Raussendorf_2007}.  One interesting proposal to create a larger two-dimensional grid using feedback and atom-photon interactions in a cavity has been put forward \cite{Pichler_2017}.  However, this approach requires coupling the emitter to a chiral waveguide, which has yet to be demonstrated experimentally with high efficiency.

In this paper, we propose an in-principle deterministic method to produce an arbitrary graph state that does not require photon-photon interactions.  Instead, the entanglement is created between emitters and then transferred onto photons through optical pumping.  This is similar to the idea presented in Ref.~\cite{PhysRevLett.105.093601}, although here, in addition to being able to create states corresponding to arbitrary graphs of any size, our scheme also allows for the entangling operations between emitters to occur \emph{after} the photon pumping.  Combined with heralding of the emission process, this allows one to perform the typically costly entangling operation only if there is a successful photon emission.  Moreover, we show that, in a precise sense, our approach allows for the maximum possible delay in transferring the entanglement from the emitters onto the photons.  Our protocol allows for arbitrary sized $n$ by $m$ clusters, given a linear array of $n$ emitters, where each emitter can be entangled with its nearest neighbors.  The operations required of the linear array are \emph{only} a single two-emitter entangling gate and a photon pumping operation.  For three-dimensional clusters, which permit \emph{fault-tolerant} universal computation, our protocol requires a two-dimensional grid of nearest-neighbor connected emitters.

This paper is organized as follows.  First, a brief overview of the graph state formalism is provided.  Second, our new framework to produce arbitrary graph states using emitters, assuming entangling gates are available between emitters, is presented.  Then, we discuss concrete realizations of \emph{parallelized} two- and three-dimensional cluster state generation protocols, which require a number of emitters that is proportional to the width of the cluster, entangling gates only between nearest-neighbors, and a time proportional only to the \emph{length} of the cluster.  We close with an assessment of realizations in ion traps, quantum dots, and nitrogen-vacancy color centers.

\section{Graph states from quantum emitters: general considerations}
A \emph{graph state} \cite{Hein_2004} is a simultaneous eigenstate of the stabilizer operators
\begin{equation}
K^{(a)}_G=X^{(a)}\prod_{b\textnormal{\scriptsize\ adjacent to }a}Z^{(b)},\label{eqn:stabilizer}
\end{equation}
with eigenvalues equal to $1$.  Here, $G$ is a graph in which each vertex represents a qubit (with logical states $\ket{0}$ and $\ket{1}$), and edges represent entanglement between the connected qubits.  There is one stabilizer for each vertex, $a$, of the graph, and it involves the application of Pauli operations to each of its neighboring vertices, $b$.  Every graph $G$ can thus be used to define a particular multi-qubit state.  The graph provides a compact and pictorial way to represent certain highly entangled states of many qubits.  Graph states can also be defined in a constructive way by first preparing each qubit in the state $\ket{+}=(\ket{0}+\ket{1})/\sqrt{2}$ and then applying a two-qubit controlled-$Z$ gate between each pair of qubits connected by an edge:
\begin{equation}
\ket{G} = \left(\prod_{b\textnormal{\scriptsize\ adjacent\ }a} \CZ_{ab}\right) \bigotimes_{c\in V} \ket{+}_c,\label{eqn:graph-cz}
\end{equation}
where $V$ are the vertices of the graph, each representing a qubit of the graph state.  The $\CZ$'s commute with each other, so the order in which they are performed is immaterial.  While this description suggests a natural way to create a graph state, such an approach will not work with, e.g., photons because there is no efficient, deterministic method to directly entangle pre-existing photons.

\begin{figure}
\centering
\includegraphics[width=5cm]{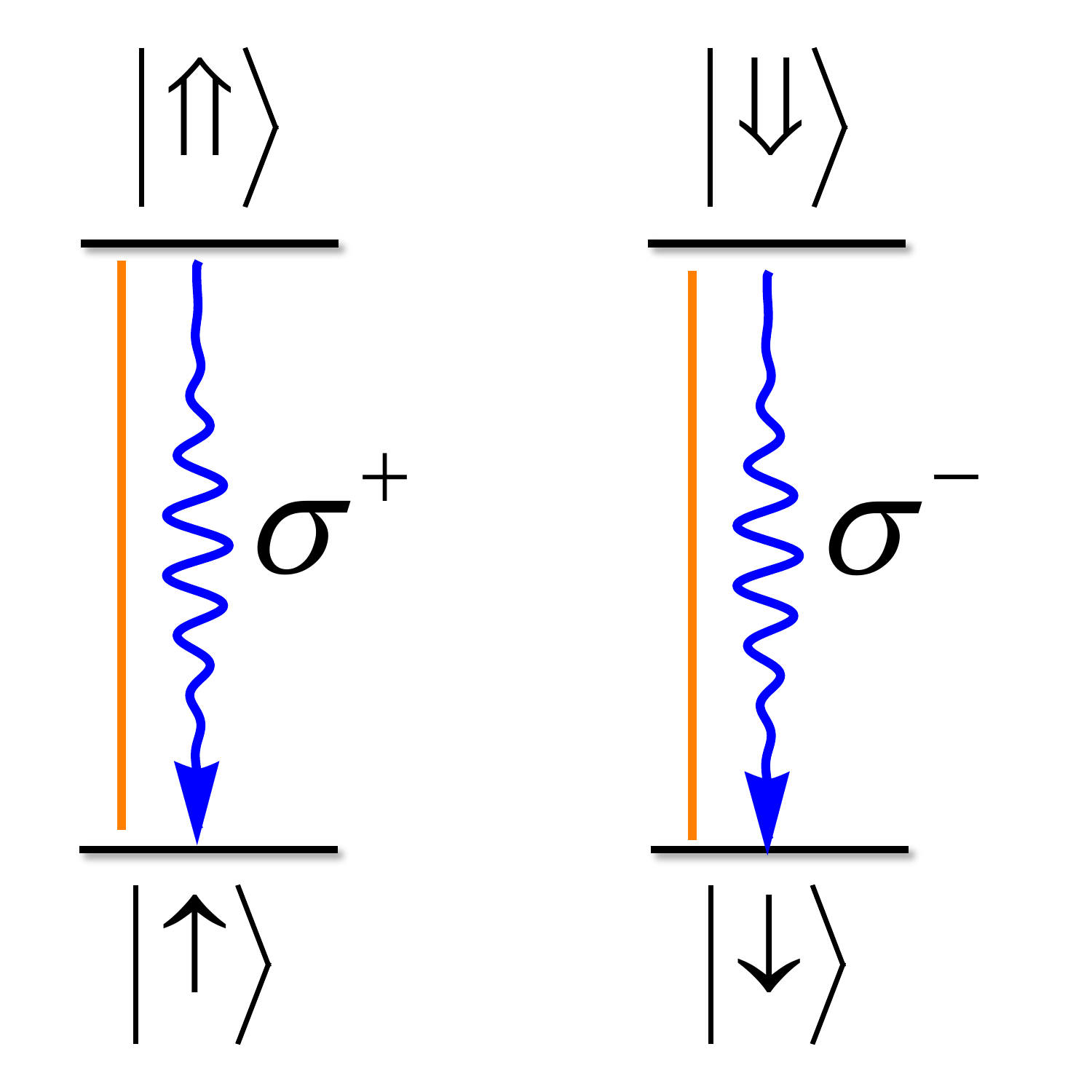}
\caption{Energy level structure in a quantum dot emitter.  The ground state electron with $J_z=+1/2$ ($\ket{\uparrow}$) and excited $J_z=+3/2$ ($\ket{\Uparrow}$) ``trion'' states shown at left are degenerate with the opposite angular momentum states at right.  Selection rules prohibit cross-excitation or decay.  Both ground states can be driven coherently up to their respective excited states by linearly polarized light (orange lines), leading to spontaneous emission from the dot (blue arrows), producing a single photon with polarization entangled with the spin degree of freedom of the quantum dot.  \label{fig:qd-levelstructure}}
\end{figure}

\begin{figure}
\centering
\includegraphics{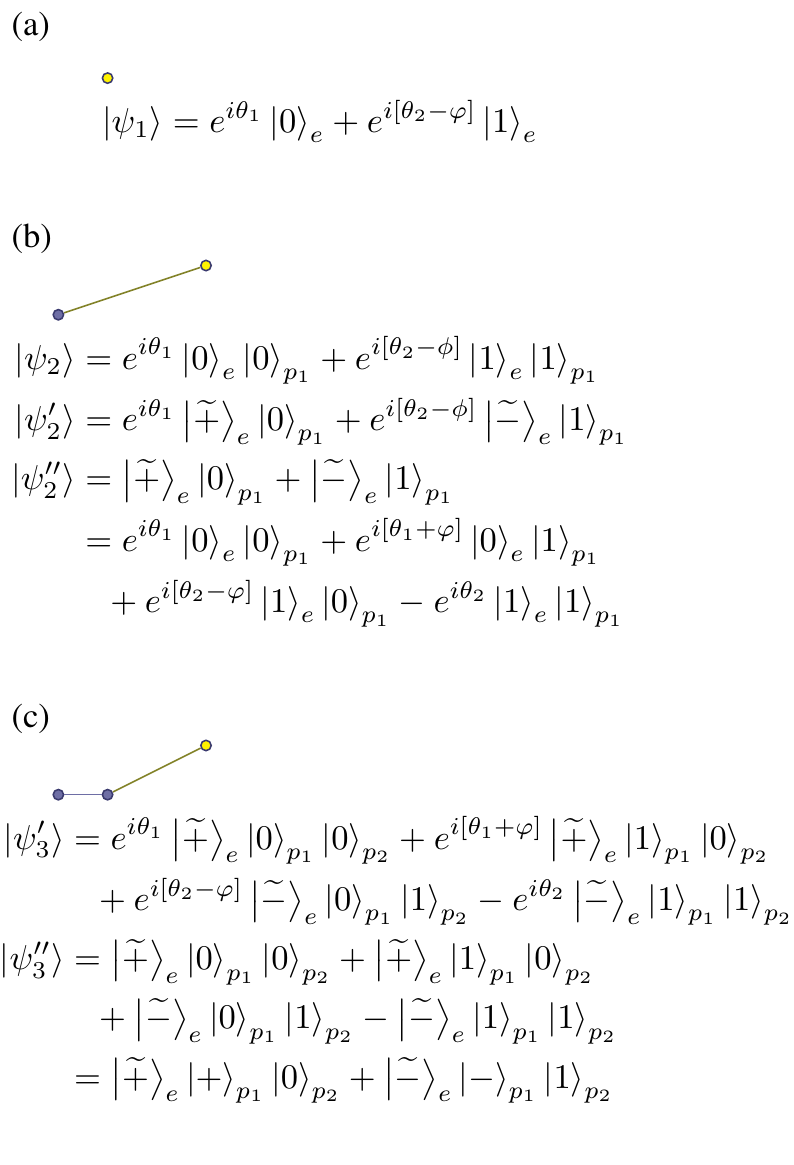}\\
\caption{Illustration of the 1D Lindner-Rudolph cluster state generation procedure.  (a) A single emitter is initialized into the state $\ket{0}$, and then $H'$ acts to produce $\ket{\psi_1}$.  A proper Hadamard $H$ would have $\theta_1=\theta_2=\varphi=0$, but we will see here that this is not necessary.  (b) The photon pumping action creates $\ket{\psi_2}$, and then $H'$ is applied to produce $\ket{\psi'_2}$.  Local phase operations on $p_1$ can remove the phases to create $\ket{\psi_2''}$, which is a graph state, up to a basis change (on the emitter only).  (Here, $H'\ket{0}=\ket{\widetilde+}$ and $H'\ket{1}=\ket{\widetilde-}$ and $\ket{\pm} = (\ket{0}\pm\ket{1})/\sqrt{2}$.)  (c) Subsequent pumping and action of $H'$ produces $\ket{\psi_3'}$.  Local phase operations on qubits $1$ and $2$ correct the phases of $H'$ in $\ket{\psi_3''}$, which is again a graph state, (again, up to a local basis change on the emitter only).  \label{fig:graph-pump}}
\end{figure}

We will focus on producing such a state using physically reasonable operations, in a way that avoids the necessity to implement entangling operations directly between photons or between photons and emitters.  We assume that the level structure and selection rules of the emitter are such that each of its two ground states optically couples to its own excited state, and that the two transitions have opposite polarization selection rules (i.e., the emitter possesses an ``II'' level system).  A prototypical example is a self-assembled quantum dot that confines a single electron spin (see \figref{fig:qd-levelstructure}).  The degenerate ground state spin manifold is mapped to the logical qubit states: $\ket{\uparrow}\to\ket{0}$ and $\ket{\downarrow}\to\ket{1}$.  In this case, if the emitter is initially in a generic superposition state, then optical excitation with linearly polarized light followed by spontaneous emission will produce a photon that is entangled with the emitter:
\begin{equation}
\alpha\ket{0}_e+\beta\ket{1}_e \to \alpha\ket{0}_e\ket{0}_{p_1}+\beta\ket{1}_e\ket{1}_{p_1},
\end{equation}
where the first ket corresponds to the emitter, and the second to the newly emitted photon.  Thus, when the emitter is prepared in a generic superposition before the pumping process, it will be entangled with the emitted photon.  Maximal entanglement will occur for an equal superposition, $|\alpha|=|\beta|$.  Such a state can be initialized by first preparing the emitter in one of the basis states and then applying a Hadamard-like gate on it,
\begin{equation}
H'=\frac{1}{\sqrt{2}}\begin{cmatrix} e^{i\theta_1} & e^{i[\theta_1+\varphi]} \\ e^{i[\theta_2-\varphi]}&-e^{i\theta_2}\end{cmatrix}.  \label{eqn:generalized-H}
\end{equation}
The Lindner-Rudolph protocol for generating 1D cluster states \cite{PhysRevLett.103.113602} amounts to alternating between applications of $H'$ and optical pumping, as is illustrated in \figref{fig:graph-pump}.  The graph can be used to define a state using either \eqnref{eqn:stabilizer} or \eqnref{eqn:graph-cz}.  Each cycle in the figure entangles a new photon to the graph.  The protocol works for any $H'$ of the form shown in \eqnref{eqn:generalized-H}, although in the remainder of this paper, we will focus on the case where $H'=H$ is a proper Hadamard gate, corresponding to $\theta_1=\theta_2=\varphi=0$.

\begin{figure}
\centering
\includegraphics{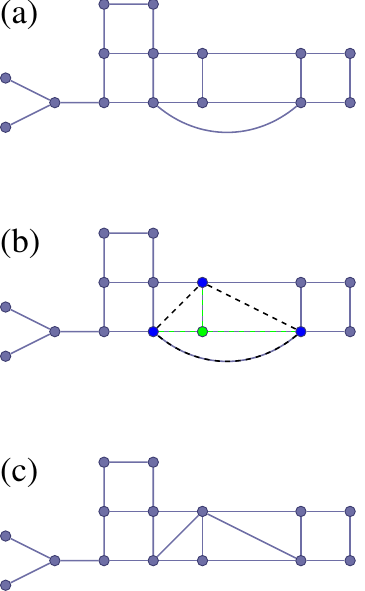}\\
\caption{A \emph{local complementation}, corresponding to $X$ ($Z$) rotations by $\pi/2$ ($-\pi/2$) on the green (blue) highlighted qubit(s).  (a) An initial graph state.  (b) All possible edges among the neighbors of the green qubit for the graph in (a) are presented as dashed black lines.  (c) Those edges are ``toggled'' in the sense that the existing ones are removed, and non-existing ones are created.  Physically, it may be more convenient to track all local complementations, and only apply the final, overall unitary at the end of the graph state creation process.  However, for expository purposes, we will apply them in sequence with other operations such as pumping and measurements as they are needed.  \label{fig:graph-LC}}
\end{figure}

The fact that repeating the pump-Hadamard cycle produces a 1D photonic cluster state can be seen from the following inductive argument.  The base case is just the emitter: $\ket{\psi_1}=\ket{+}$.  Assume, inductively, that the emitter is already entangled with $n$ photons in a 1D cluster state $\ket{\psi_{n+1}}$ with the emitter at one end.  We can write the state as $\ket{\psi_{n+1}}=\ket{0}_e\ket{\psi_{n}}+\ket{1}_e\ket{\bar\psi_{n}}$, where $\ket{\bar\psi_n}=Z_n\ket{\psi_n}$ and $Z_n$ is the Pauli-$Z$ operator acting on the $(n-1)$th photon emitted (and we will validate this inductive hypothesis momentarily).  Now consider pumping the emitter and then applying $H$ to the emitter:
\begin{eqnarray}
\ket{0}_e\ket{\psi_n} + \ket{1}_e\ket{\bar\psi_n} \to
\ket{0}_e\ket{\psi_n}\ket{0}_{p_n} + \ket{1}_e\ket{\bar\psi_n}\ket{1}_{p_n} \to\nonumber\\
\hspace{14em}\ket{+}_e\ket{\psi_n}\ket{0}_{p_n} + \ket{-}_e\ket{\bar\psi_n}\ket{1}_{p_n}.
\label{eqn:lindner-rudolph-form1}
\end{eqnarray}
Pumping the emitter $e$ yields the first transformation, with the emitted photon labelled by $p_n$---it is the $n$th photon in the cluster.  The second transformation is the action of $H$ on the emitter.  The final expression can be expanded as
\begin{equation}
\ket{0}_e \overbrace{\left( \ket{\psi_n}\ket{0}_{p_n}+\ket{\bar\psi_n}\ket{1}_{p_n}\right)}^{\ket{\psi_{n+1}}} + \ket{1}_e \overbrace{\left( \ket{\psi_n}\ket{0}_{p_n}-\ket{\bar\psi_n}\ket{1}_{p_n}\right)}^{\ket{\bar\psi_{n+1}}}.  \label{eqn:lindner-rudolph-form2}
\end{equation}
This is the Schmidt decomposition of a state in which the emitter is now entangled with $n+1$ photons.  More precisely, the multi-photon state multiplying $\ket{0}_e$ in this decomposition is essentially just the original cluster state $\ket{\psi_{n+1}}$ we started with, but where the emitter has been replaced by the newly emitted photon $p_n$.  Moreover, \eqnref{eqn:lindner-rudolph-form2} is the state we would get if we started with $(\ket{\psi_{n}}\ket{0}_p+\ket{\bar\psi_n}\ket{1}_p)\ket{+}_e$ and then applied a $\CZ$ gate between photon $p$ and the emitter.  This implies that it is the 1D cluster state $\ket{\psi_{n+2}}$ for $n+1$ photons and the emitter, with the emitter attached at one end.  Thus repeating the pump-Hadamard cycle $n$ times produces a 1D cluster state involving $n$ photons.

Now suppose that we replace the state $\ket{\psi_n}$ (and $\ket{\bar\psi_n}$) in \eqnsref{eqn:lindner-rudolph-form1}{eqn:lindner-rudolph-form2} with a general graph state $\ket{\Psi}$ (and $\ket{\bar\Psi}$) that may involve multiple photons \emph{and} other emitters with which the emitter is entangled.  The pumping and Hadamard operations still act in the same way, leading to the conclusion that these two processes together essentially insert a new photon into the graph in the place of the emitter, while the emitter gets pushed out from the graph and is only connected to the new photon.  This is one way to understand why entangled emitters produce entangled photons, as originally shown in \cite{PhysRevLett.105.093601}.  The fact that this mechanism holds for a general graph features prominently in the protocols we present here.

It is important to note that the edges of the graph do not uniquely determine the entanglement structure of the corresponding state.  In particular, edges can appear or disappear under the application of purely local unitary operations.  The exact class of graph states that are related to each other by local unitary (LU) operations is not fully understood \cite{Hein_2004,Van_den_Nest_2004}.  The rich \cite{quant-ph/0504166v1,Tsimakuridze_2017,Zeng_2007,1805.05968}, but smaller \cite{0709.1266v2,Tsimakuridze_2017} set of \emph{local Clifford} (LC) operations can be understood in graph-theoretic terms as ``local complementations'' \cite{Hein_2004,Van_den_Nest_2005} of the graph (see \figref{fig:graph-LC}).  These operations are also very useful for the protocols we introduce below; they describe an equivalence class of graphs that can be accessed from one another by local unitary operations.  In the present context of all-photonic graph states, this corresponds to waveplate operations.  In the following section, we describe an experimentally meaningful coarsening of these equivalence classes: we additionally allow $\CZ$ entangling gates on preferred vertices (i.e., the quantum emitters).

\section{General scheme for emitting graph states}
\begin{figure}
\centering
\includegraphics{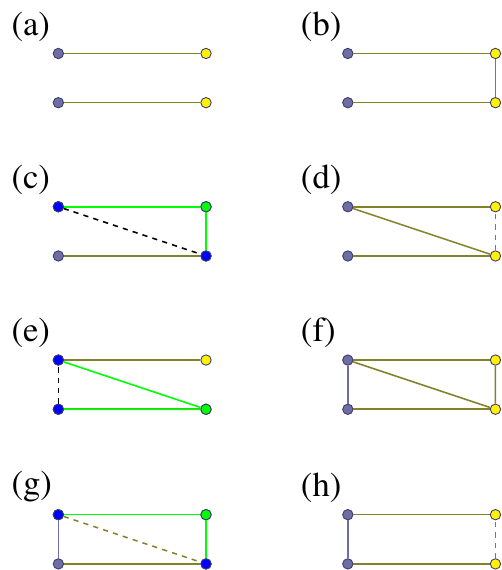}\\
\caption{(a) Two emitters are pumped creating a graph state, with photons in blue, and emitters in yellow.  Entanglement edges are represented as lines connecting vertices.  (b) A $\CZ$ gate is performed between the emitters.  (c) A local complementation is performed on emitter 1, realized by an $X$ rotation on emitter 1 (colored green), and $Z$ rotations on the neighboring qubits (colored blue, connected by green highlighted edges), inducing a new entanglement edge (dashed and colored black).  (d) A $\CZ$ gate is performed between the emitters.  Notice that, for the simplified graph shown here, a local complementation on the first photon would be sufficient to remove the entanglement between the emitters.  However, if the two photons were part of a larger graph state, the local complementation would disturb the rest of the graph.  (e) A local complementation is performed on emitter 2.  (f) A $\CZ$ gate is performed on the emitters.  (g) A local complementation is performed on emitter 1.  (h) A $\CZ$ gate is performed on the emitters.  These steps themselves do not need to each be performed separately: \eqnref{eqn:entangler} compiles all the emitter operations (b)-(h) into one gate, $\mathfrak{g}$; the local unitary operations commute with the operations on the emitters, and can be performed at any time.  \label{fig:primitive}}
\end{figure}

\begin{figure}
\centering
\includegraphics{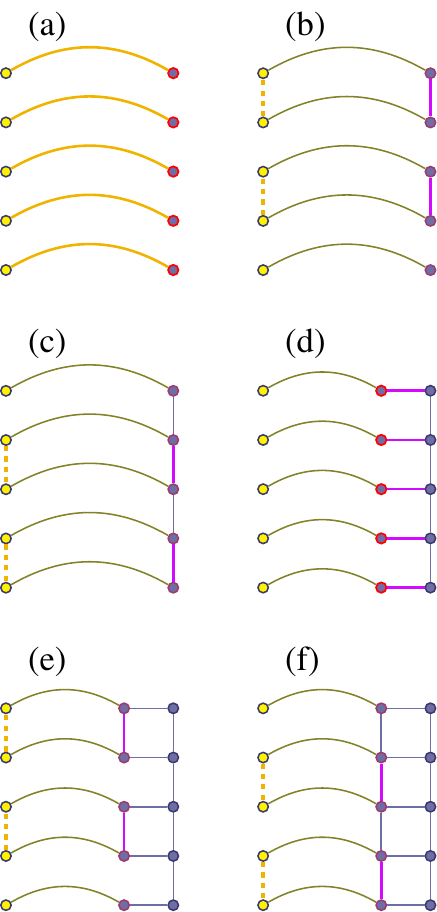}
\caption{Parallelized $n\times m$ 2D cluster state generation protocol.  The protocol requires $n$ emitters that can be entangled with their nearest neighbors via $\CZ$ gates.  The coloring is as in \figref{fig:primitive}.  (a) Initially, each emitter is pumped once, creating $n$ entangled emitter-photon Bell states.  ($n=5$ in this illustration).  (b and c) $\mathfrak{g}$ gates are performed on odd (and then even) pairs of neighboring emitters, inducing entanglement on the newly created photons.  (d) One cycle of the Lindner-Rudolph 1D cluster state generation protocol is performed.  (e-g) $\mathfrak{g}$ is again applied on odd (and then even) pairs of emitters, again inducing entanglement in the existing graph state.  Steps (d) through (f) are repeated to create a full $n\times m$ all-photonic cluster state.  Notice that this scheme naturally generalizes to $k$-dimensional cluster states, printing $(k-1)$-dimensional ``sheets'' of photons at a time, but requiring as many emitters as there are photons in each sheet.
\label{fig:2dcluster-par}}
\end{figure}

The emitter-based graph state generation scheme described in the previous section involves first creating entanglement between emitters and then pumping the emitters to create entangled photons.  However, instead of creating the entanglement and then pumping, which requires \emph{all} entanglement edges to have been created before performing the pumping action, the entanglement edges can be produced \emph{after} emission of the photons, as is illustrated in detail in \figref{fig:primitive}.  This may lead to simplifications in the experimental implementation of such schemes.  For instance, if the pumping process admits some mechanism for heralding the successful emission of a photon, then one could choose whether or not to implement subsequent entangling gates based on this information.  Because all local unitary operations on the photons commute with all operations on the emitters (except for the pumping), the operations on the emitters described in \figref{fig:primitive} can be collected into a single entangling operation:
\begin{equation}
\mathfrak{g}=
\CZ U^{(1)}_{1}U^{(2)}_{2}  \CZ U_2^{(1)} \CZ U_2^{(2)} U_1^{(1)} \CZ \propto i (X\otimes\mathbf{1}) + \mathbf{1}\otimes X,\label{eqn:entangler}
\end{equation}
where $U^{(1)}$ and $U^{(2)}$ are $X$ and $Z$ rotations (by $\pi/2$ and $-\pi/2$), and the subscripts 1 and 2 indicate which emitter the operator is being applied to.  We emphasize that $\mathfrak{g}$ adds precisely one edge between a pair of photons that have already been emitted, irrespective of other qubits' entanglement with the two photons.  There remains some freedom in the choice of $U^{(1)}$ and $U^{(2)}$ (and some choices may be preferred for a given type of emitter), but regardless of this choice, $\mathfrak{g}$ is locally equivalent to a maximally entangling gate.  We take $U^{(1)}$ and $U^{(2)}$ as rotations about the $X$ and $Z$ axes for concreteness here.  Other combinations of gates might allow for better parallelization, e.g., the scheme of \figref{fig:2dcluster-par} might avoid the final $\CZ$, because it could in principle be performed in parallel.  Finally, notice that
\begin{equation}
i\SWAP \mathfrak{g}^\dag \SWAP = \mathfrak{g}, \label{eqn:swapped-entangler}
\end{equation}
which explains that our choice of which emitter to perform the first local complementation on (in \figref{fig:primitive}) is immaterial.

\begin{figure*}[t]
\centering
\includegraphics{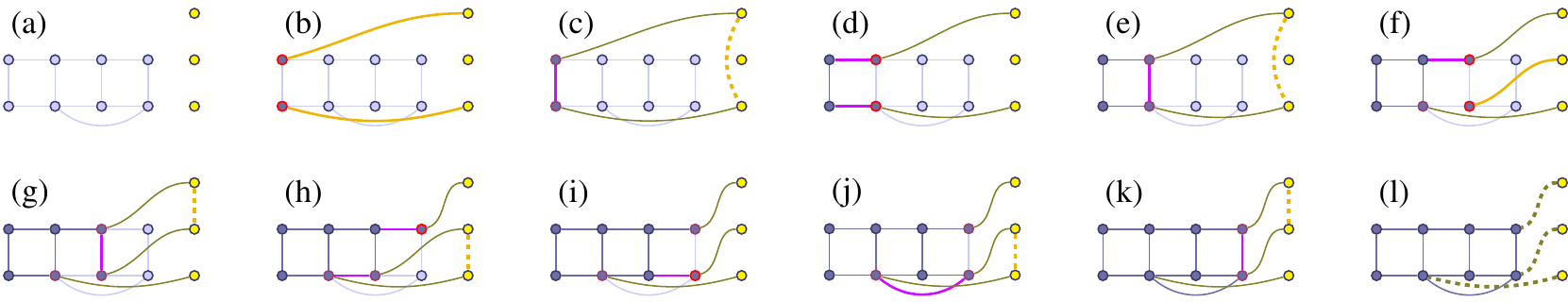}
\caption{Photon-by-photon graph generation.  (a) An example of a target graph state, with yet-to-be created photons and entanglement edges colored faintly.  For this particular target state, we assume we have at our disposal three emitters in a linear array.  These are shown on the right as yellow dots.  (b) The first and third emitters are pumped (the first emitter is the one at the top), producing one entangled photon each (shown in bright red), and forming two emitter-photon entangled Bell pairs (yellow lines).  (c) The $\mathfrak{g}$ gate is performed on the first and third emitters, inducing an entanglement edge on the recently emitted photons (in heavier, purple color).  The pale red highlight of the two photons indicates they are not yet fully entangled with all desired photons and therefore remain connected to the emitters.  (d) One cycle of the Lindner-Rudolph 1D cluster protocol \cite{PhysRevLett.103.113602} is performed on emitters 1 and 3, creating two new photons and a pair of respective entanglement edges.  The first two photons are now properly entangled with all appropriate photons, and are shown without red highlighting.  (e) Step (c) is repeated, creating another edge.  (f) The 1D cluster protocol is performed on emitter 1, and emitter 2 is pumped once.  Emitter 3 is idle.  (g and h) $\mathfrak{g}$ is performed on emitters 1 and 2 (respectively, 2 and 3), inducing new entanglement edges.  (i) The 1D protocol is performed on emitter 2, producing the last photon.  (j and k) $\mathfrak{g}$ is performed on emitters 2 and 3 (1 and 2, respectively), producing the final entanglement edges.  (l) The emitters are disconnected from the graph state, either by pumping and measuring an additional photon, or by direct $z$ measurement of the emitters.  \label{fig:graph-layers}}
\end{figure*}

We now have in place all the basic ingredients we need to describe our general protocol for creating arbitrary graph states.  The full protocol is illustrated in \figref{fig:graph-layers}, and is, in summary,
\begin{enumerate}
\item[1.] The emitters are initialized to
$\ket{+}=(\ket{0}+\ket{1})/\sqrt{2}$.
\item[2.] Emitters are pumped and $H'$ is applied for each photon in the desired ``active'' layer.
\item[3.] Entanglement edges are created using optimized $\mathfrak{g}$, until at least one photon has every desired entanglement edge (and possibly one extra, attached to the emitter).
\item[4.] The emitter attached to that photon is pumped, and $H'$ is applied.
\item[5.] If the newly emitted photon is not part of the target state (i.e., if the entanglement edge to the emitter was ultimately unneeded), it is measured in the $Z$ basis, disconnecting the emitter from the graph state.  If another photon is desired in the active layer, the emitter is pumped and $H'$ applied once again.
\item[6.] Steps 3-5 are repeated until all photons and edges are created.
\end{enumerate}

\begin{figure}
\centering
\includegraphics{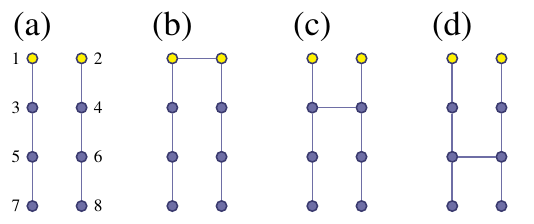}\\
\caption{Four graph states, inequivalent under local unitary (or local Clifford) operations.  (a) Two three-photon 1D cluster states are entangled with two respective emitters.  The qubits are labeled, and photons are blue while the emitters (1 and 2) are yellow.  (b) As before, with the emitters entangled with each other.  (c) As shown in \figref{fig:primitive}, by applying a combination of local unitary operations and entangling operations \emph{between only the emitters}, the ``rung'' of the ladder can be shifted down one position.  (d) The rung of the ladder \emph{cannot} be moved down an additional step by only local unitary operations and \emph{arbitrary} operations between emitters.  This would require introducing a stabilizer $X_6 Z_4 Z_8 Z_5$.  It is clear that any nontrivial Clifford gate on photon 5, simultaneously \emph{trivially} affecting photon 7, must include the stabilizer $Z_5X_3Z_1$.  However, $X_3$ cannot be removed from the stabilizer; neither local Clifford conjugation nor multiplication by another stabilizer can convert it to the identity.  This extends straightforwardly to generic local unitaries (see \ref{appendix:proof}).  \label{fig:lu-equiv}}
\end{figure}

For step 3 to be performed, the emitters must have sufficient connectivity to produce the appropriate entanglement edges.  That connectivity, and the number of emitters required, depends on the graph state and the order that the photons are produced.  This can be understood in terms of the number of the ``active photons'' which are directly linked to an emitter via a graph edge and waiting to be connected to additional photons (e.g., the photons outlined in red in \figref{fig:graph-layers}).  The maximal number of photons in this layer determines the number of required emitters, and the connectivity between photons in that layer corresponds to the connectivity required between emitters.  In particular, each photon is kept directly entangled with an emitter (in the graph state sense through an entanglement edge) until all entanglement edges associated with that photon have been constructed.  Among the photons still ``attached'' in this sense to the emitters, (i.e., the active layer), entanglement can be produced between emitters and transferred to the existing photons via $\mathfrak{g}$ and appropriate local operations on the emitted photons.  Once the emitters are pumped, the ability to move entanglement edges of the graph state around is significantly limited.

To illustrate why photons need to remain attached to emitters until all their connections are created, and to understand better why a single $\CZ$ gate is not in general sufficient to create an entanglement edge between two photons, consider the graph state shown in \figref{fig:lu-equiv}.  Could a single $\CZ$, creating a single entanglement edge between emitters, be sufficient to produce an entanglement edge between photons 5 and 6, i.e, can we get from (b) to (d) in \figref{fig:lu-equiv} by only local waveplate operations? The answer is no.  For graphs of $8$ or fewer qubits, the equivalence class of graphs related by local unitary operations is the same as those related by local Clifford unitaries, or equivalently local complementations \cite{Cabello_2009,Hein_2004,Hein_2004,Van_den_Nest_2004}.  There are efficient methods to check if two graphs are related by local complementation \cite{Bouchet_1991}, allowing us to check that the four graphs of \figref{fig:lu-equiv} are inequivalent.  However, if $\CZ$ gates are allowed on the emitters, the equivalence classes generically coarsen.  We have shown, in the construction of $\mathfrak{g}$, a sequence of such gates that relate (a) and (c), so they are in the same coarsened equivalence class.  The remaining question is ``Can (d) be obtained from (a) by only local waveplate operations and \emph{arbitrary} unitary operations on the emitters?'' The answer to that question is no.  As explained in \figref{fig:lu-equiv}, for (a) and (d) to belong in the same coarsened class, it would have to be possible to create the stabilizer $X_6 Z_4 Z_8 Z_5$ (which stabilizes state (d)) from products of stabilizers of (a), local Clifford unitaries, and arbitrary unitaries on the emitters.  Because this stabilizer commutes with both $Z_3X_1$ and $Z_5X_3Z_1$, the action on $3$ must be trivial and so this is not possible.  See \ref{appendix:proof} for a full proof.

At this point, it is worth making a few comments about the efficiency and resource requirements of our protocol.  First of all, the protocol requires the gate $\mathfrak{g}$ to be performed for each edge in the graph.  However, the number of edges in a graph is not invariant under local unitary operations.  A carefully chosen local unitarily-equivalent graph state may be significantly faster to produce.  Whether or not there exists an equivalent graph with significantly fewer edges likely depends strongly on the particular graph in question, and therefore so do the performance and resource requirements of the protocol.  Finding optimal graphs that are local unitary-equivalent to a graph of interest remains an active area of research \cite{Dahlberg_2018}.  Furthermore, allowing for the measurement of a photon during the generation of the graph state may also significantly reduce the emitter requirements.  While the three main operations required on the emitters (Hadamard gates, the $\mathfrak{g}$ gate, and the emitter pumping) are certainly no simple experimental feat, they constitute a subset of the Clifford group.  By the Gottesman-Knill theorem \cite{Gottesman1998}, quantum circuits built from these gates can be efficiently simulated on a classical computer.  In other words, the gate set requirements for our protocol are less demanding than those needed for universal quantum computation.  Another important feature of our protocol is that the photons are emitted first and the entanglement happens afterward.  Experimentally, such a capability could be important in cases where the emitters do not always produce photons.  Our protocol then allows the entangling gate, which is typically the costliest component of any quantum information protocol, to be performed only if photons are produced.  Let us consider that the photon emission is probabilistic.  If a heralding mechanism is available, then the emitter could be pumped repeatedly until a photon emission is heralded, and the entangling gate could only be applied upon the successful emission of a photon.  Such a heralding mechanism could originate from a cascade transition, which is possible in quantum dots where multi-excitonic states are common, in which case the first emitted photon would be measured in the transverse basis, projecting it out from the state without affecting the entanglement of the second emitted photon.  Another, more technically demanding, option would be a QND detection of the emitted photons.  Let us also remark on the unsuccessful events; one may think that such events destroy the cluster state, as they can indeed open `holes' in the state, where the line connecting photons to the emitters is cut.  Fortunately, it has been shown that even if there are holes, so long as the percolation threshold is met, the cluster state can still be used for universal quantum computing \cite{Browne_2008}.  Because the unsuccessful events will not all happen at the same layer, for sufficiently high emission probability and number of emitters, a useful (percolated) cluster state can be generated.

We now show how this protocol can be used to create a 2D cluster state of arbitrary size $n\times m$.  The step-by-step procedure is illustrated in \figref{fig:2dcluster-par}.  The protocol assumes that we have $n$ emitters (where $n$ is the smaller dimension of the graph) arranged in a linear array, and that we have the capability to perform entangling $\mathfrak{g}$ gates between nearest neighbors.  The protocol essentially amounts to alternating between applying multiple $\mathfrak{g}$ gates between adjacent emitters in parallel and performing pump-Hadamard operations.  One subtlety is that we must apply the $\mathfrak{g}$ gates in two layers each time to avoid having a single emitter undergo entangling operations with two of its neighbors simultaneously.  The fact that we can perform all the $\mathfrak{g}$ gates in each layer at the same time allows for a significant savings in the overall time required to implement the protocol.  This scheme naturally generalizes to $d$-dimensional cluster states; a $w^{d-1}\times m$ cluster state would require $w^{d-1}$ emitters with connectivity between nearest neighbors, and time proportional to $m\times 2(d-1)+\textnormal{constant}$ (the factor of $2(d-1)$ comes from the need to produce each edge in the active layer).

\begin{figure}
\centering
\includegraphics{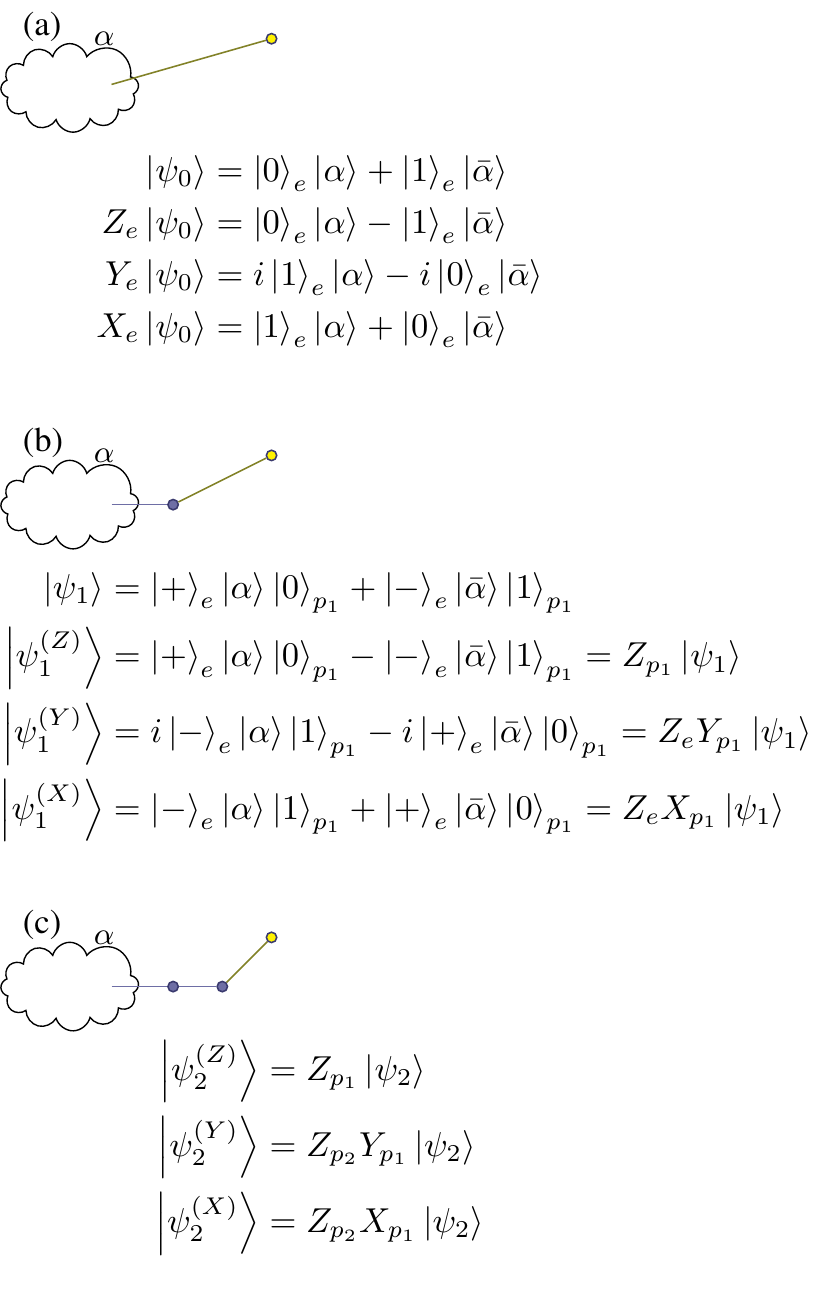}
\caption{Propagation of errors from emitters to photons.  (a) A quantum emitter (yellow dot) entangled with an existing graph state (visualized as a cloud $\alpha$), $\ket{\psi_0}$.  Three Pauli errors are described.  (b) The pumping-Hadamard procedure is performed on the emitter, ideally producing $\ket{\psi_1}$, but if one of the aforementioned Pauli errors occurred, instead produces $\ket{\psi_1^{X}},\ket{\psi_1^{Y}},\ket{\psi_1^{Z}}$.  These states are related to $\ket{\psi_1}$ by errors on the emitter or photon.  Notice there is some freedom in the choice of this representation of error.  (c) The pumping-Hadamard procedure is repeated.  Notice that any of the three original Pauli errors can be represented as errors on at most two emitted photons.  \label{fig:error}}
\end{figure}
We close out this section with a discussion of errors and their propagation during the generation of the graph state.  Even the highest quality systems will inevitably develop at least some error during their operation---whether it be from photon loss, decoherence of the emitters, or non-ideal emitter gates.  As in other graph state generation protocols \cite{PhysRevLett.103.113602,PhysRevLett.105.093601,Russo_2018}, the ``pumping'' step of the graph state generation tends to reduce the severity of the error on the emitter, at the expense of producing corrupted photons.  The other stages, such as the $U^{(2)}U^{(1)}$ operation of the 1D cluster case or $\mathfrak{g}$ of the present protocol, simply transform the type of error.  For example, if we commute a Pauli error past $\mathfrak{g}$ from right to left, we obtain a residual error operator that can be interpreted as the error that remains on the emitter after $\mathfrak{g}$ is applied.  \eqnref{eqn:swapped-entangler} allows us to only consider errors on the first qubit.  To wit:
\begin{equation}
\mathfrak{g}X_1 =X_1\mathfrak{g},
\end{equation}
which is obvious, because $X$ commutes with itself.  The remaining Pauli errors, on the other hand, have more complicated commutation properties:
\begin{equation}
\mathfrak{g}Z_1 =-Y_1X_2\mathfrak{g},
\end{equation}
and
\begin{equation}
\mathfrak{g}Y_1 =Z_1X_2\mathfrak{g}.
\end{equation}
Therefore, pre-existing $Y$ or $Z$ errors induce an $X$ error on emitters that are entangled via $\mathfrak{g}$.  The $X$ errors, however, will not propagate any further, localizing the region of damage in the emitter layer, at that step.

During the pumping process, as in the 1D cluster state case, errors are passed into the photonic states: $Z$ errors induce an $X$ error on the new photon; $X$ errors induce a $Z$ photon error on the second most recently emitted photon; and finally, $Y$ errors induce an $X$ error on the emitted photon and a $Z$ error on the second most recently emitted photon \cite{Russo_2018}.  As additional local operations are performed on the photons (i.e., to implement the local complementations), these will transform further (see \figref{fig:error}).

Any error existing just before the entangling $\mathfrak{g}$ gate is applied will propagate no further than the nearest-neighbor emitters.  After the subsequent pumping step, the quantum state of the emitter is put in a form where all errors are in the photons---while maintaining the unaffected emitters entanglement with the photonic graph state.  Thus, if an emitter is entangled with $n$ other emitters, at most $n+2$ corrupted photons will be produced.  The error localizes, though errors on neighboring sites are \emph{not} uncorrelated: fault tolerant algorithms will need to be adjusted to handle the pattern of errors created.

\begin{figure}
\centering
\includegraphics[width=5cm]{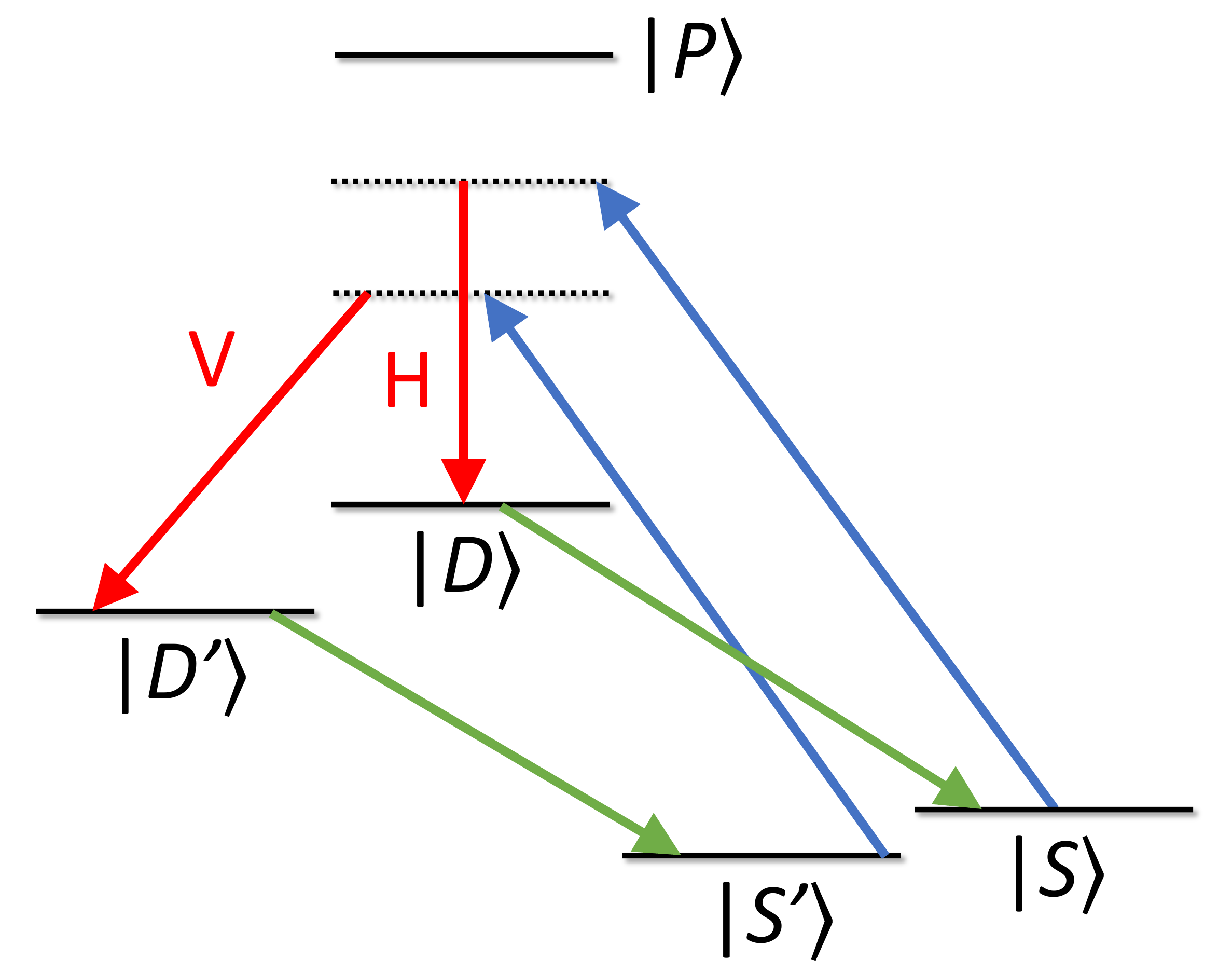}
\caption{Raman driving from two ground $S$ states, split by a Zeeman field (blue lines) detuned from a $\ket{P}$ state.  Emitted photons (red lines) would, in free space, be circularly or $\pi$-polarized, but in a cavity are projected to horizontal and vertical polarizations as the system relaxes into two metastable states $\ket{D}$ and $\ket{D'}$ (maximizing scattering strength).  Coherent manipulations (green lines) return the system from the metastable $D_{5/2}$-manifold to the $S_{1/2}$ Zeeman-split ground states.  \label{fig:raman-ions}}
\end{figure}

\begin{figure}
\centering
\includegraphics[width=5cm]{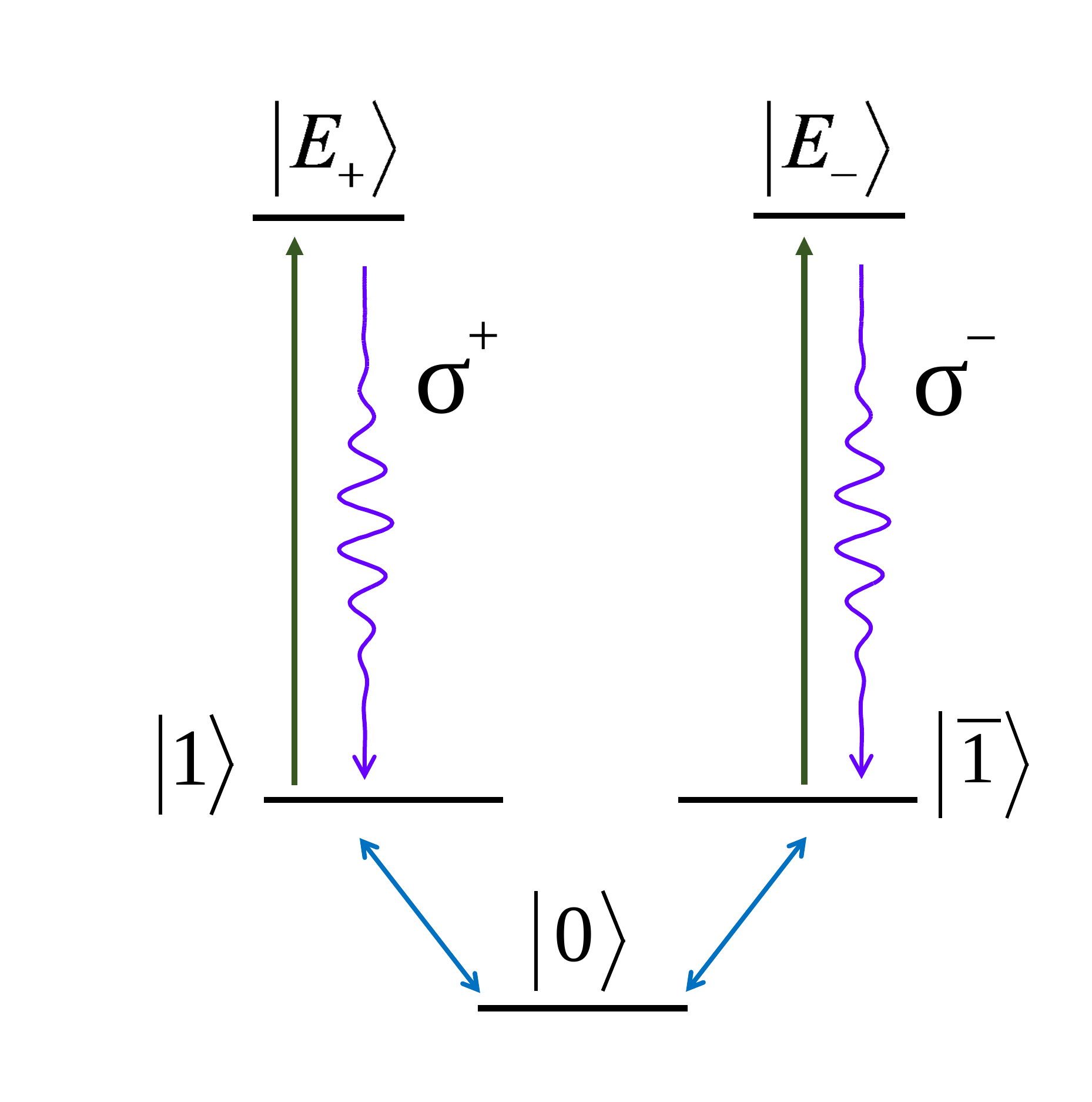}
\caption{Energy level structure in a negatively-charged nitrogen-vacancy color center in diamond.  Microwave driving (blue lines) initializes the system into the $|J_z|=1$ subspace (comprising the degenerate states $\ket{1}$ and $\ket{\bar 1}$ with $J_z=+1$ and $J_z=-1$, respectively) and enables generic unitary gates.  These optically active states can be excited (green arrows) coherently into the $|J_z|=2$ manifold (comprising $\ket{E_\pm}$), from which spontaneous emission yields circularly polarized photons (purple arrows).
\label{fig:nv-leveldiagram}}
\end{figure}

\section{Experimental Realizations}

Trapped ions exhibit remarkably useful properties for the purposes of quantum information applications, including stability on the order of minutes \cite{Wang_2017}, single-photon-matter quantum interfaces \cite{Wilk_2007,Stute_2012,Stute_2012_2,Sterk_2012,Steiner_2013}, high-fidelity universal gates \cite{Gaebler_2016,Ballance_2016}, and entanglement in quantum registers, involving as many as 20 trapped ions \cite{Monz_2011,Friis_2018}.  Critically, the required integration of these techniques is already underway \cite{Herold_2016}.  Widely recognized as a promising platform for quantum computation \cite{Lekitsch_2017}, trapped ions possess all the essential capabilities necessary to implement our graph state generation protocols.

The energy levels of a single trapped ion, e.g., ${}^{40}\mathrm{Ca}^+$, in a cavity, are shown schematically in \figref{fig:raman-ions}.  A magnetic field orthogonal to the cavity axis induces a Zeeman splitting in each of the $S$, $P$, and $D$ manifolds, enabling sophisticated state-specific Raman scattering (see \cite{Stute_2012,Stute_2012_2}).  In particular, states from the metastable $D_{5/2}$ manifold are Raman-resonantly excited from the Zeeman-split ground states $\ket{S}$ and $\ket{S'}$ with $m=+1/2,-1/2$, respectively, by a monochromatic laser pulse (compare with the dichromatic pulse of \cite{Stute_2012_2}).  The states $D$ and $D'$ are carefully selected to maximize the overall Raman scattering strength \cite{Stute_2012}.  The system is initialized (by cooling) to the $\ket{S}$ and $\ket{S'}$ states, and then excited via dichromatic pumping detuned from the $\ket{P}$ (with $m=-3/2$) as in \cite{Stute_2012_2}.  This architecture has been realized in \cite{Stute_2012}, and allows for fast ($\sim1\units{\mu s}$) single-qubit operations and photon generation ($\sim20\units{\mu s}$), with relatively long coherence times ($\sim250\units{\mu s}$) limited primarily by slow instabilities of the magnetic field strength.  Pumping multiple ions in the same cavity requires each ion be precisely positioned at the antinodes of the standing wave---and this condition can be stabilized by the presence of multiple ions \cite{Begley_2016}.  The other ions in the trap in the hyperfine manifold will be unresponsive to the scattered (pumped) photon.

\begin{figure}
\centering
\includegraphics{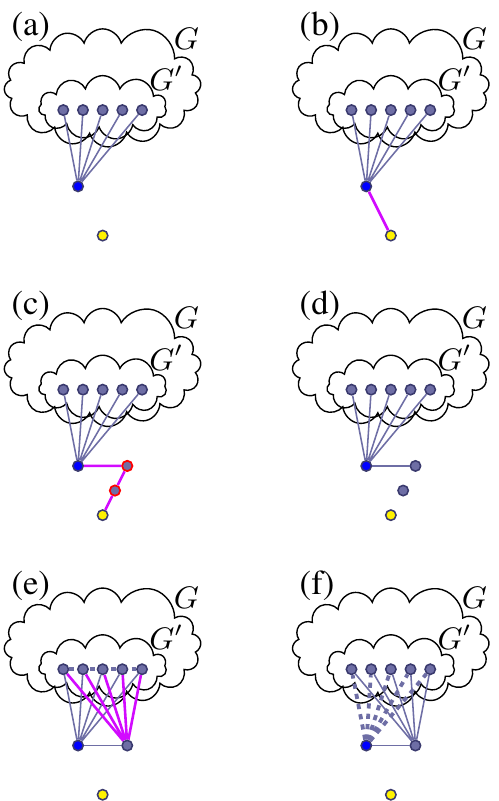}
\caption{Alternative pumping approach with nuclear spins and NV centers.  (a) A nuclear spin (blue) is entangled with a graph state $G$ (but only directly entangled with $G'$), and the NV center (yellow) is initialized, but disentangled.  (b) The NV center is entangled via a $\CZ$ gate.  (c) The NV center is pumped \emph{twice}.  (d) The second photon is measured in the $Z$ basis, disconnecting the NV center from the graph.  (e) A local complementation is performed on the nuclear spin, generating connectivity with the new photon.  (f) A local complementation is performed on the new photon, removing the direct connection between the nuclear spin and $\ket{G}$.  Notice that this second round resets the connectivity within the subgraph $G'$.  \label{fig:NV-pump}}
\end{figure}

Fast ($\sim 10\units{\mu s}$), high-fidelity ($\gtrsim 99\%$) two-qubit gates have already been realized in cavities \cite{Ballance_2016}, suggesting that the $\mathfrak{g}$ gates required for our protocol could be implemented with high precision.  In fact, well-controlled entangling operations driven by laser fields have very recently been demonstrated with up to $20$ ions \cite{Friis_2018}.  Because our approach only requires entangling a pair of neighboring ions at a time, we anticipate that the main experimental challenge in implementing our protocol, at least for cluster states of size $\lesssim 20$, lies in integrating the various operations together.

This architecture also holds promise for scaling up to the much larger cluster state dimensions required for actual quantum computations.  The parallelized protocol depicted in \figref{fig:2dcluster-par}, for example, can be compatible with setups involving many trapped ions distributed across several different traps.  To apply the entangling gate $\mathfrak{g}$ on emitters in different traps, a pair of optically active ancilla ions, one in each trap, are employed.  The ancillae are first entangled by a heralded, but not necessarily deterministic, approach, such as \cite{Casabone_2013}, where a successful joint measurement of photons in a Bell state entangles the emitting ancillae.  Notice that in the event of failure, this entangling operation can be retried without affecting the rest of the graph state by simply restarting the entangling step.  Intra-trap gates between the ancilla ion and the desired emitter, along with subsequent ancilla measurement, are sufficient to create the maximally entangling $\mathfrak{g}$.  Moreover, not every attempt at generating entanglement needs to be successful: after just two attempts, the probability of generating an entanglement bond is at least $3/4$, which is greater than the percolation threshold for two-dimensional (and three-dimensional) square lattices \cite{Stauffer_1979,quantum_percolation_2009}.  Given a percolating subset of the cluster state, it is possible to efficiently perform generic quantum computation \cite{Gross_2006,Kieling_2007}.  Alternatively, there is interest in shuttling ions between traps within separate cavities, which could then be directly entangled by intra-trap techniques \cite{1367-2630-18-10-103028,Nizamani_2011,Bowler_2012,Wright_2013,Hucul_2014}.  Provided that inter-trap operations between unrelated pairs can be performed in parallel, the scaling of such a device would take time proportional to the width $n$ of the cluster state, rather than the total number of emitted photons.  For the fault-tolerant three-dimensional case, this significantly improves the theoretical performance.  The higher-dimensional case requires a two, rather than one, dimensional array of ion traps, which has been demonstrated \cite{Kumph_2011}.

Another promising candidate architecture for our scheme is self-assembled quantum dots, with fast, optically controlled interactions \cite{PhysRevB.74.205415,Press_Nature08,Greilich_NP09,Carter_2013}, comparatively long lifetimes \cite{PhysRevLett.116.033603}, and fast photonic pumping \cite{PhysRevLett.103.113602,Senellart_NatNano17,PhysRevB.90.155303,Birowosuto_2012,Kelaita_2016}.  The Lindner-Rudolph protocol \cite{PhysRevLett.103.113602} has in fact been used to experimentally demonstrate the creation of a linear five-photon cluster state from a quantum dot emitter \cite{Schwartz_2016}.  The energy level diagram for a negatively charged quantum dot is shown in \figref{fig:qd-levelstructure}.  Here, $\ket{\uparrow}$ and $\ket{\downarrow}$ are the spin states of the confined electron.  Each of these ground states is coupled via circularly polarized light to a $J_z=\pm 3/2$ trion, denoted by $\ket{\Uparrow}$ and $\ket{\Downarrow}$, respectively.  (A trion is a bound state of an exciton and an electron.)  Linearly polarized light drives both transitions such that if the electron spin is prepared in a superposition, then coherent excitation followed by spontaneous emission (with time scale $T_1\sim$ ns) produces a photon whose polarization is entangled with the electron spin:
\begin{equation}
\alpha\ket{\uparrow}+\beta\ket{\downarrow} \to \alpha\ket{\Uparrow} + \beta\ket{\Downarrow} \to \alpha\ket{\uparrow}\ket{R}+\beta\ket{\downarrow}\ket{L}.\nonumber\label{eqn:polarized-drive}\\
\end{equation}
Because our scheme requires only nearest-neighbor entangling gates, an array of stacked quantum dots \cite{PhysRevLett.94.057402,Stinaff636} could potentially be used to generate a 2D cluster state.  Creating larger arrays and achieving high photon collection efficiency are likely among the most challenging aspects of this approach.

A third candidate architecture that we will discuss is negatively charged nitrogen-vacancy (NV) color centers in diamond or silicon carbide.  The relevant energy levels of an NV center in diamond are shown in \figref{fig:nv-leveldiagram} \cite{Economou_2016}.  Optical pumping initializes the NV into the state $\ket{0}$, and microwave driving subsequently brings the system into the metastable $|J_z|=1$ qubit subspace, spanned by $\ket{1},\ket{\bar{1}}$.  Light-matter entanglement can be achieved very much analogously to \eqnref{eqn:polarized-drive}: linearly-polarized coherent light drives an excitation that then decays, emitting a polarized photon ($T_1\sim 10\units{ns}$).  While this decay time is slower than that of the quantum dot, the coherence time of these centers is also longer---up to a millisecond at room temperature in isotopically purified samples \cite{Balasubramanian_2009} that has been extended to nearly a second using dynamical decoupling \cite{Bar_Gill_2013}, with continuing effort to further improve this via, e.g., quantum error correction \cite{Cramer_NatComm2016}.  Directly entangling these nitrogen vacancy centers is challenging, but (non-deterministic) heralded entanglement generation has been demonstrated \cite{Hensen_2015}.  A failure to entangle the NV centers would require a full restart of the graph state generation, making such an approach unsuitable.  However, by using an ancillary quantum state, such as a nearby nuclear spin (e.g., ${}^{14}C$), repeated attempts to produce heralded entanglement can be made without having to start the whole graph state generation process over.  Then, even with a weak ($\gtrsim 50\units{kHz}$) hyperfine coupling between the nuclear spin and the NV-center, microwave manipulations can be used to implement fast ($\sim 30\units{\mu s}$) entangling gates between them \cite{Russo_2018}.  For this architecture, we propose a variant of our protocol where the nuclear spins play the role of the ``active'' layer in \figref{fig:graph-layers}---a single nuclear spin per emitter is required.  This allows non-deterministic operations to be performed on the NV center without disturbing the existing graph state.

Both the pumping and $\mathfrak{g}$ operation are slightly different in this version of the protocol.  We first describe the pumping operation.  Because the nuclear spins are not optically active, the spin-photon interface must be realized via the NV-center.  The most straightforward approach is to perform a full $\textsc{swap}$ gate (equivalent to two entangling gates), pump the NV center, and then perform another full $\textsc{swap}$ (to return the quantum state to the stable nuclear spin).  Two full $\textsc{swap}$ operations may take a prohibitively long amount of time.  An alternative approach may be faster: (1) entangle the NV-center with the nuclear spin; (2) pump the NV-center (interspersed with $H'$ operations) \emph{twice}; (3) measure the second photon to disentangle the emitter; and finally (4) perform two local complementations (one on the nuclear spin, and one on the first emitted photon), producing the desired connectivity, where the new photon has taken the place of the nuclear spin in the graph (see \figref{fig:NV-pump}).  Again, as with $\mathfrak{g}$, all the local operations commute with each other, so all operations on the nuclear spins and NV-center can be compiled into a single pulse.  Although both procedures are more complex than simple microwave driving, the limited exposure of the long-lived nuclear spin to the noisier NV-center may improve overall fidelity.  Notice the trade off between performing two entangling operations (i.e., a swap) vs two pumping operations.

Next, we describe realizations of the $\mathfrak{g}$ operation.  Similar to the ion trap case, we first obtain heralded entanglement of the NV centers by entanglement swapping \emph{while they are disentangled from the nuclear spin}.  Next, entangle the NV centers with their respective nuclear spins deterministically with a microwave pulse \cite{Russo_2018}.  We are guided by the results of \figref{fig:lu-equiv}: using only local unitaries, the entanglement cannot be transferred from the NV centers to the nuclear spin.  Instead, we complete the protocol by pumping each NV-center twice, measuring the first two photons in the $Y$ basis, and the second two photons in the $Z$;  the $Z$ measurements isolate the nuclear and photonic quantum state from the NV center, and the $Y$ measurements collapse the entanglement edges directly to the ancilla.  Compare this scheme to \figref{fig:primitive} where, instead of using a second entangling operation, two $Y$ measurements are made.  As in the ion trap architecture, because the entanglement is heralded, a single failed attempt to entangle the NV-centers can be retried without restarting the whole graph state generation procedure..  Similarly, the $\mathfrak{g}$ does not have to succeed, because incomplete cluster states (with probability of missing entanglement edges below a percolation threshold) are still sufficient for universal quantum computing.

NV-centers suffer from low out-coupling efficiencies, though this is conventionally combatted by coupling the center to a cavity \cite{PhysRevX.7.031040}, which has the added benefit of reducing emission out of the zero-phonon line by the Purcell effect.

\section{Outlook}

We presented an approach to generate large-scale cluster states suitable for measurement based quantum computing.  The approach requires a one- or two-dimensional array of emitters to produce a two- or three-dimensional cluster state.  The time it takes to produce these states scales with the linear dimension of the state, provided the emitters can be controlled individually.  Quantum errors in the emitters localize in the sense that they only affect a limited number of emitted photons, allowing for the possibility of fault-tolerant quantum computation.  Additionally, the protocol can be generalized to arbitrary graph states, with the required number of emitters scaling with the generalized ``width'' of the graph state, and the time with the ``length,'' again as long as the emitters can be manipulated independently.  Because the protocol only requires nearest-neighbor entanglement, ion traps, quantum dots, and nitrogen-vacancy color centers are all promising platforms for carrying out the protocol.

\section*{Acknowledgements}

This research was supported by the NSF (Grant No. 1741656).  We thank Dr.  Kenneth Brown, Natalie Brown, and Paul Hilaire for helpful discussions.

\appendix
\section{Inequivalnce of Graph States\label{appendix:proof}}
\begin{figure}
\centering
\includegraphics{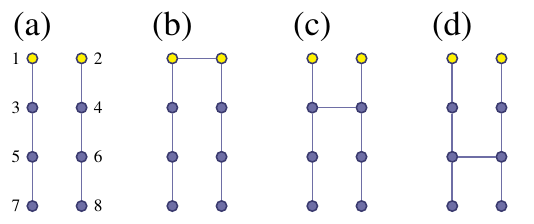}\\
\caption{Duplicate of \figref{fig:lu-equiv}, rephrased in the language of the proof: Four graph states, inequivalent under local unitary (or local Clifford) operations.  (a) Two, four-qubit 1D cluster states, with two qubits in the control manifold, colored yellow.  (b)  The control manifold is entangled here.  This is trivially $LU_\mathcal{C}$ equivalent to the first graph.  (c) As shown in \figref{fig:primitive}, this is also $LU_\mathcal{C}$ equivalent to the first two graph states.  (d) An $LU_\mathcal{C}$ inequivalent graph state, per the proof.  \label{fig:lu-equiv-dup}}
\end{figure}

In this appendix, we will demonstrate that subgraphs (a) and (d) of \figref{fig:lu-equiv} (reproduced in \figref{fig:lu-equiv-dup} for ease of reference) cannot be obtained from the other via any local unitary in combination with arbitrary unitary action on the emitters.  We begin by establishing notation.  Consider a simple graph, $G$ with vertices $\mathcal{V}[G]$ (corresponding to physical qubits) and with an adjacency matrix $N$ ($N_{ij}=1$ if $i$ and $j$ are adjacent, and $0$ otherwise).  Given a subset of physical qubits, the subgraph $G_1$ induced by this subset has vertices $\mathcal{V}[G_1]\subset \mathcal{V}[G]$.  The adjacency $N^1_{ij}$ is simply the restriction of $N_{ij}$ to $i,j\in\mathcal{V}[G_1]$.

Definition: Let a set of vertices $\mathcal{C}\subset \mathcal{V}$ be given (assume $\mathcal{V}[G]=\mathcal{V}[G']=\mathcal{V}$.  Two graphs $G$ and $G'$ are said to be $\mathrm{LU}_\mathcal{C}$-equivalent if there exists a product $U$ of locally unitary operators on $\mathcal{V}$ and $V$ a unitary operator on $\mathcal{C}$, such that
\[ UV \ket{G} = \ket{G'}.\]
The set $\mathcal{C}$ is called the \emph{control manifold}.

Recall that, if $G$ is a graph with vertices $\mathcal{V}$ and an adjacency matrix $N$, the graph state
\begin{equation}
\ket{G}=\sum_{\vec a \in 2^{V[G]}} (-1)^{\sum_{i<j} a_iN^0_{i,j}a_j} \ket{\vec a}.
\end{equation}
The sum over $2^{V[G]}$ gives all assignments of $\ket{0}$ or $\ket{1}$ to each vertex in $G$.   The sum in the exponent over $i<j$, $i,j\in \mathcal{V}[G_1]$ may be omitted when there can be no confusion for the range of $i$ and $j$.

The above expression can be decomposed in terms of subgraphs.  Consider a graph state $\ket{G}$ with subgraphs $G_0$ and $G_1$, induced by vertices $\mathcal{V}[G_0]$ and $\mathcal{V}[G_1]$, which together constitute a partition of $\mathcal{V}$.  Using an implied summation over $i$ and $j$ in the appropriate domains,
\begin{eqnarray}
\ket{G} &= \sum_{\stackrel{\vec a\in 2^{V[G_0]}}{\vec b\in 2^{V[G_1]}}} (-1)^{a_iN^{0}_{ij}a_j+a_iN^{0,1}_{ij}b_j+b_iN^{1}_{ij}b_j} \ket{\vec a,\vec b}\nonumber \\
&= \sum_{\stackrel{\vec a\in 2^{V[G_0]}}{\vec b\in 2^{V[G_1]}}} (-1)^{a_iN^{0}_{ij}a_j} \left(\prod_j Z_{(1,j)}^{a_iN^{0,1}_{ij}}\right) (-1)^{b_iN^1_{ij} b_j} \ket{\vec a,\vec b}\nonumber \\
&= \sum_{\vec a\in 2^{V[G_0]}} (-1)^{a_iN^{0}_{ij}a_j} \ket{\vec a} \otimes \prod_{ij} Z_{(1,j)}^{a_iN_{ij}^{0,1}}\ket{G_1}.
\end{eqnarray}
$N^{0}$ and $N^{1}$ are the adjacency matrices on $G_0$ and $G_1$, respectively. $N^{0,1}_{ij}$  is the adjacency matrix on $G$, restricted to $i\in \mathcal{V}[G_0]$, and $j\in \mathcal{V}[G_1]$.

Now, we precisely state our claim: Consider the case of a pair of 1D clusters of length $\geq 4$ as in \figref{fig:lu-equiv}(a), where qubits $1$ and $2$ constitute the control manifold.  We will show by contradiction that this graph is not LU${}_\mathcal{C}$ equivalent when links are added three steps into the chain, as in subfigure (d).  To wit, let $\ket{G}$ and $\ket{G'}$ be the graph states of \figref{fig:lu-equiv-dup}(a) and (d), respectively.  Assume we have $U$ and $V$ as in the definition of LU${}_\mathcal{C}$, ($UV\ket{G}=\ket{G'}$).  Our argument will find that $V$ does not entangle qubits $4$ and $8$, which is a contradiction because the linear clusters would remain disentangled.  Alternatively, this may be viewed as a demonstration that adding the control manifold does \emph{not} coarsen the equivalence class, without directly showing that subfigures (a) and (d) are in different LU classes.  This second form generalizes naturally to any supergraph of \figref{fig:lu-equiv}.

Proof: Because $U$ is a local unitary, we can decompose it into parts $U_i$ that act individually on each qubit (other than $1$ and $2$, whose effect is assumed to be absorbed into $V$):
\[U=\prod_{n\neq 1,2} U_i.\]
Next, define a subgraph $G_0$ induced by the vertices $4$ through $8$.  $\ket{G}$ can be decomposed as follows,
\begin{eqnarray}
 \ket{G} =&
   &\ket{0}_3\ket{+}_1\ket{0}_2 \ket{G_0}\nonumber \\
  &+&\ket{0}_3\ket{+}_1\ket{1}_2 Z_4 \ket{G_0}\nonumber \\
  &+&\ket{1}_3\ket{-}_1\ket{0}_2 Z_5 \ket{G_0}\nonumber \\
  &+&\ket{1}_3\ket{-}_1\ket{1}_2 Z_5Z_4 \ket{G_0}.\label{eqn:3-decomp-G}
\end{eqnarray}
$Z_3X_1$ is a stabilizer of $UV\ket{G}$ because $UV\ket{G}=\CZ_{5,6}\ket{G}$:
\begin{equation}
\overbrace{U_3^\dag Z_3 U_3}^{\tilde Z_3} \overbrace{V^\dag X_1 V}^{\tilde X_1}\ket{G} = \ket{G}.
\end{equation}
Each of the 4 terms in \eqnref{eqn:3-decomp-G}, i.e., $\ket{G_0}$, $Z_4 \ket{G_0}$, $Z_5 \ket{G_0}$, and $Z_5 Z_4 \ket{G_0}$, are orthogonal for any graph state $\ket{G_0}$ (because the $Z_i$ anticommute with exactly one graph stabilizer).  Applying projectors onto those states gives, respectively
\begin{eqnarray*}
\left(\tilde Z_3\ket{0}_3\right)\otimes
 \left(\tilde X_1\ket{+}_1\ket{0}_2\right)
 &= \ket{0}_3 \otimes \left( \ket{+}_1\ket{0}_2 \right)\\
\left(\tilde Z_3\ket{0}_3\right)\otimes
 \left(\tilde X_1\ket{+}_1\ket{1}_2\right)
 &= \ket{0}_3 \otimes \left( \ket{+}_1\ket{1}_2 \right)\\
\left(\tilde Z_3\ket{1}_3\right)\otimes
 \left(\tilde X_1\ket{-}_1\ket{0}_2\right)
 &= \ket{1}_3 \otimes \left( \ket{-}_1\ket{0}_2 \right)\\
\left(\tilde Z_3\ket{1}_3\right)\otimes
 \left(\tilde X_1\ket{-}_1\ket{1}_2\right)
 &= \ket{1}_3 \otimes \left( \ket{-}_1\ket{1}_2 \right).
\end{eqnarray*}
We conclude that $\tilde Z_3$ is diagonal on $\ket{0},\ket{1}$, i.e. $\tilde Z_3=ce^{i\alpha Z_3}$, and $\tilde X_1$ is diagonal on $\ket{+}\ket{0},\ket{+}\ket{1},\ket{-}\ket{1},\ket{-}\ket{1}$.  Moreover, $\tilde X_1$ acts trivially on qubit $2$, and its phase on $\ket{1}$ must exactly cancel that of $\tilde Z_3$, so that $\tilde X_1=\bar ce^{-i\alpha X_1}$.

Using an Euler decomposition $U_3=e^{i\zeta Z_3}e^{i \eta X_3}e^{i\xi Z_3}$ on \[U_3^\dag Z_3 U_3 = \tilde Z_3 =c e^{i \alpha Z_3},\]
we get
\[e^{-i\xi Z_3}e^{-i \eta X_3}e^{-i\zeta Z_3} Z_3 e^{i\zeta Z_3}e^{i \eta X_3}e^{i\xi Z_3}= ce^{i\alpha Z_3},\]
or
\[e^{-i \eta X_3} Z_3 e^{i \eta X_3}= ce^{i\alpha Z_3},\]
i.e., we can assume $\zeta=\xi=0$.  Because $Z_3=-ie^{i(\pi/2) Z_3}$,
\[Z_3^\dag e^{-i \eta X_3} Z_3 e^{i \eta X_3}= -ice^{i(\alpha+\pi/2) Z_3}.\]
or
\[e^{-i \eta Y_3} e^{i \eta X_3}= -ice^{i(\alpha+\pi/2) Z_3}.\]
It follows that $\eta=0$ or $\eta=\pi$ (because acting on $\ket{0}$, must give at most a phase), and we get
\[1 = -ice^{i(\alpha+\pi/2) Z_3}.\]
Therefore, $c=\pm i$, $\alpha=\mp\pi/2$.  It follows that $\tilde X_1=X_1$ and $U_3=I$.

This approach is repeated for the stabilizer $Z_5X_3Z_1$ (notice, critically, that $Z_5$ commutes with $\CZ_{5,6}$). Again,
\[\overbrace{U_5^\dag Z_5 U_5}^{\tilde Z_5} \overbrace{U_3^\dag X_3 U_3}^{\tilde X_3}\overbrace{V^\dag Z_1 V}^{\tilde Z_1}\ket{G} = \ket{G}.\]
Except this time, we have not come unarmed: $U_3=I$, so $\tilde X_3=X_3$.  $G_0$ is now the subgraph with qubits $4$, and $6$ through $8$, making the decomposition significantly larger:
\begin{eqnarray}
 \ket{G} =
 & &\ket{0}_5\ket{0}_3\ket{+}_1 \left[\ket{0}_2+\ket{1}_2 Z_4\right]\ket{G_0}\nonumber\\
 &+&\ket{1}_5\ket{0}_3\ket{+}_1 \left[\ket{0}_2+\ket{1}_2 Z_4\right]Z_7\ket{G_0}\nonumber\\
 &+&\ket{0}_5\ket{1}_3\ket{-}_1 \left[\ket{0}_2+\ket{1}_2 Z_4\right]\ket{G_0}\nonumber\\
 &-&\ket{1}_5\ket{1}_3\ket{-}_1 \left[\ket{0}_2+\ket{1}_2 Z_4\right]Z_7\ket{G_0}\nonumber\\
\hphantom{\ket{G}}=
  & &\ket{0}_5(\ket{0}_3\ket{+}_1+\ket{1}_3\ket{-}_1) \ket{0}_2\ket{G_0}\nonumber\\
  &+&\ket{0}_5(\ket{0}_3\ket{+}_1+\ket{1}_3\ket{-}_1) \ket{1}_2Z_4\ket{G_0}\nonumber\\
  &+&\ket{1}_5(\ket{0}_3\ket{+}_1-\ket{1}_3\ket{-}_1) \ket{0}_2Z_7\ket{G_0}\nonumber\\ &+&\ket{1}_5(\ket{0}_3\ket{+}_1-\ket{1}_3\ket{-}_1) \ket{1}_2Z_7Z_4\ket{G_0}.
\end{eqnarray}
The last stage separates based on orthogonal states of $G_0$. Acting $\tilde Z_5X_3\tilde Z_1$:
\begin{eqnarray}
 \tilde Z_5X_3\tilde Z_1 \ket{G} = \hspace{-6.5em}&&\nonumber\\
  &&
   \tilde Z_5 \ket{0}_5(\ket{1}_3\tilde Z_1\ket{+}_1
   +\ket{0}_3\tilde Z_1 \ket{-}_1) \ket{0}_2\ket{G_0}\nonumber\\
  &+&\tilde Z_5\ket{0}_5(\ket{1}_3\tilde Z_1\ket{+}_1
   +\ket{0}_3\tilde Z_1\ket{-}_1) \ket{1}_2 Z_4\ket{G_0}\nonumber\\
  &+&\tilde Z_5\ket{1}_5(\ket{1}_3\tilde Z_1\ket{+}_1
   -\ket{0}_3\tilde Z_1\ket{-}_1) \ket{0}_2 Z_7\ket{G_0}\nonumber\\
  &+&\tilde Z_5\ket{1}_5(\ket{1}_3\tilde Z_1\ket{+}_1
   -\ket{0}_3\tilde Z_1\ket{-}_1) \ket{1}_2 Z_7Z_4\ket{G_0}.
\end{eqnarray}
Notice that, as before, $\tilde Z_5$ is diagonal on $\ket{0}_5,\ket{1}_5$. It follows that $\tilde Z_1=Z_1$. Finally:
\begin{equation}
V^\dag Y_1 V = V^\dag X_1Z_1 V = V^\dag X_1 VV^\dag Z_1 V = \tilde X_1 \tilde Z_1 = X_1Z_1 = Y_1
\end{equation}

It follows that $V^\dag (A_1\otimes I_2) V=A_1\otimes I_2$ for \emph{any operator} on qubit $1$. The same argument above holds for qubit $2$, and therefore $V$ is a pure phase, and cannot entangle the chains.

\bibliography{references}
\bibliographystyle{iopart-num}

\end{document}